\documentclass[11pt,a4paper]{article}
\usepackage[utf8]{inputenc}
\usepackage[T1]{fontenc}
\usepackage{lmodern}

\usepackage{graphicx}
\usepackage[font={footnotesize, sf}, labelfont=bf]{caption} 

\usepackage{amsmath, amssymb, mathtools}
\usepackage{amsthm, thmtools}
\usepackage{tikz-cd}
\usepackage{siunitx}

\usepackage{csquotes}

\usepackage{enumerate}

\usepackage{geometry}
\usepackage{setspace}
\usepackage{hyperref}
\usepackage[english, capitalise, noabbrev]{cleveref}
\hypersetup{
	colorlinks,
	linkcolor = {red!50!blue},
	citecolor = {red!50!blue},
	urlcolor = {red!50!blue},
	linktoc = page
}

\usepackage[style=alphabetic, citestyle=alphabetic]{biblatex}
\usepackage[english]{babel}
\newcommand{\qedtriangle}{\ensuremath{\triangle}}
\newcommand{\qedtriangledown}{\ensuremath{\bigtriangledown}}
\allowdisplaybreaks

\declaretheoremstyle[spaceabove=6pt, spacebelow=6pt, headfont=\bfseries,
notefont=\normalfont, notebraces={(}{)}, qed=\qedtriangle]{definition}
\declaretheoremstyle[spaceabove=6pt, spacebelow=6pt, headfont=\bfseries,
notefont=\normalfont, notebraces={(}{)}, qed=\qedtriangledown]{example}

\usepackage[bf,sf,small,pagestyles]{titlesec}
\usepackage{titling}

\titleformat{\chapter}[block]{\sffamily \bfseries \Huge}{\filleft \large Chapter \Huge \thechapter\\}{0pt}{\Huge \titlerule[1pt] \vspace{1ex} \filleft}

\usepackage{subfiles}

\newcommand{\R}{\mathbb{R}}

\renewcommand{\L}{\mathcal{L}}

\newcommand{\ud}{\,\mathrm{d}}

\renewcommand{\d}{\mathrm{d}}

\makeatletter
\newcommand*{\defeq}{\mathrel{\rlap{%
    \raisebox{0.3ex}{$\m@th\cdot$}}%
  \raisebox{-0.3ex}{$\m@th\cdot$}}%
	=
}
\makeatother

\DeclareMathOperator{\tr}{tr}
\DeclareMathOperator{\Ric}{Ric}

\title{\sffamily \bfseries A variational derivation of the field equations of an action-dependent
	Einstein-Hilbert Lagrangian}
\author{\sffamily 
 $^a$Jordi Gaset
 \thanks{\texttt{jordi.gaset@unir.ne}t\,({ORCID}:\,0000-0001-8796-3149).} ,
$^b$Arnau Mas Dorca
\thanks{\texttt{arnau.mas@physik.uni-muenchen.de}\,({ORCID}:\,0000-0003-0532-0938).}.
\\[1ex]
\normalsize\itshape\sffamily 
$^a$Escuela Superior de Ingeniería y Tecnología, Universidad Internacional de La Rioja, Spain.
\\[1ex]
\normalsize\itshape\sffamily 
$^b$ Fakultät für Physik, Ludwig-Maximilians-Universität München
}
\date{\sffamily \small June 23, 2022}

\begin{document}
\maketitle

\selectlanguage{english}

\begin{abstract}
We derive the equations of motion of an action-dependent version of the Einstein-Hilbert Lagrangian, as a specific instance of the Herglotz variational problem. Action-dependent Lagrangians lead to dissipative dynamics, which cannot be obtained with the standard method of Lagrangian field theory. First-order theories of this kind are relatively well understood, but examples of singular or higher-order action-dependent field theories are scarce. This work constitutes an example of such a theory. By casting the problem in clear geometric terms we are able to obtain a Lorentz invariant set of equations, which contrasts with previous attempts. 
\end{abstract}

\newpage
	{\small \sffamily \tableofcontents}
\newpage

\section{Background and motivation}

Most of the physical systems of interest in physics admit a description in terms of a
variational principle: physical solutions are the extrema of some functional defined on
the state of all possible evolutions of the system, called the action. If the action is
defined in terms of the integral of a Lagrangian then its extrema are precisely the
solutions to the Euler-Lagrange equations of the Lagrangian. For mechanical systems, i.e.
systems in which solutions are functions only of time, the phase space can be equipped
with a symplectic structure. From this point of view, time evolution is just the flow
generated by the Hamiltonian of the system, and one has at one's disposal all of the tools
known from symplectic mechanics: Poisson bracket, Noether's theorem, etc. The geometric
framework can be generalized to field theories using, for instance, the multisymplectic
formalism.

This description, nevertheless, excludes a large class of systems, namely dissipative
systems. In some cases the phase space for such systems is naturally equipped with a
contact structure, which can in many ways be seen as the odd dimensional analogue of a
symplectic structure (see \cite{grabowska_geometric_2022} for a more general discussion).
What in the symplectic world were conservation laws, now become dissipation laws
\cite{Gaset2020b}. Contact structures have made appearances in various fields in recent
years: reversible and non-reversible thermodynamics
\cite{Bravetti2019,Mrugala1991,Simoes2020}, quantum mechanics \cite{ciaglia_contact_2018},
statistical mechanics \cite{goto_contact_2016}, cosmology \cite{Lazo2017,Sloan} or
electromagnetism \cite{GasetMarin}. The contact framework is well understood for
mechanical systems, see
\cite{Gaset2020b,geiges_introduction_2008,Lainz2019,Leon2021,Leon2019,Leon2021a,cocontact}.
The field theory analog, multicontact geometry is under current development, see
\cite{Gaset2020, Gaset2020a,Georgieva, de_leon_multicontact_2022} for recent efforts. One
of the main challenges is a successful understanding of singular and higher-order
theories.

In this work we study one such theory, namely a dissipative version of Einstein gravity.
To circumvent the difficulties coming from the as of now not fully understood contact
formalism, we make use of the fact that contact systems, when seen from the Lagrangian
point of view, can also be formulated in terms of a variational principle, the Herglotz
variational principle. The Lagrangians that fit in this framework are called
\emph{action-dependent}. We apply variational calculus to derive the analog of the
Euler-Lagrange equations, the Herglotz equations, for this system. This is of relevance
since examples of singular or higher-order dissipative field theories are scarce in the
literature. 

The result obtained is also relevant to the study of modifications of Einstein's theory of
gravity, that would explain some observations of cosmological phenomena that do not fit
within the current picture, as well as open avenues towards the successful quantisation of
gravity. A survey of theories of this kind is in \cite{Olmo2020} and in
\cite{baker_strong_2017}. In \cite{Lazo2017,Lazo2021}, the same Lagrangian we introduce
was studied. We frame it within the broader context of the Herglotz variational principle
and dissipative theories. In this sense this work is complementary to \cite{Lazo2017}, in
terms of clarifying the geometric nature of the objects at play and writing down a set of
equations that is Lorentz invariant, as opposed to the ones originally derived. This issue
is also remedied in \cite{Lazo2021}.

The work is organised as follows: in \cref{ch:herglotz} we introduce the Herglotz
variational principle and show how it can be equivalently formulated as a constrained
optimisation problem. This allows one to use calculus of variations to derive the correct
Herglotz equations of motion, especially relevant for field theories. In
\cref{ch:einstein} we apply this language to a dissipative version of the Einstein-Hilbert
Lagrangian to derive its field equations. In \cref{ch:significance} we discuss how these
equations differ from the ones originally derived in \cite{Lazo2017} and why they are a
Lorentz invariant generalisation of them.

\section{The Herglotz variational problem}\label{ch:herglotz}

This section presents the theory of \emph{action-dependent Lagrangians}. The main appeal of
this formalism is that it allows for the description of non-conservative systems in terms of a
variational principle, which is in general not possible with standard Lagrangian
mechanics. The problem of finding the stationary paths of the action given by a Lagrangian of this sort is known as the Herglotz problem \cite{Herglotz_1930}. The main difficulty of this variational
problem is that, as opposed to the standard variational problem of Lagrangian mechanics,
it is an implicit optimisation problem. 

The phase space of an action-dependent Lagrangian theory can be equipped with a contact structure. Hence, from the Hamiltonian point of view, contact geometry is the natural framework in which to describe dissipative dynamics. This is well-understood for mechanics, but not mature enough for field theories, in particular for  second-order theories like the Hilbert-Einstein Lagrangian. This work will focus on the Lagrangian picture and variational methods.

There are several ways of deriving the equations of motion of a Herglotz variational principle in mechanics. The original version, defines a functional on trajectories in terms of the solution to a differential equation determined by the trajectory. We refer to this as the implicit approach. Alternatively, one can implement the action-dependence as a non-holonomic constraint on a standard variational problem defined on a larger configuration space and use standard variational methods. There are two distinct ways of implementing non-holonomic constraints, which are referred to as the vakonomic method and the non-holonomic method. For the Herglotz principle in mechanical systems, they are shown to be equivalent in \cite{de_leon_constrained_2021}, in the sense that they lead to the same equations.

The implicit approach to the Herglotz principle for field theories presents a difficulty because the differential equation that needs to be solved is now a partial differential equation. Nevertheless, this has been successfully done for a particular class of first-order field theories  in \cite{Georgieva, Lazo2018}. Here we instead follow \cite{de_leon_constrained_2021} and use the constrained approach.

This is, to the authors' knowledge, the first time that the equations of motion for a second-order action-dependent field theory have been derived. Hence, although the resulting equations are physically and geometrically sound, they cannot be compared to other results of this sort. This is relevant because, in general, the vakonomic and non-holonomic methods are not equivalent, and only one of them leads to the desired result \cite{Gra2003}. This is clarified by the authors in collaboration with M. Lainz and X. Rivas in \cite{GLMR-2022}. The key result is that when the action dependence is closed then both methods are equivalent.

We now present the Herglotz principle for mechanical systems and first order-field theories, and show how the Herglotz equations are derived using the vakonomic method.

\subsection{The Herglotz variational problem as constrained optimisation}
An action-dependent Lagrangian is defined on the configuration space of the non-dissipative theory expanded with an extra degree of freedom. This additional degree of freedom is on-shell interpreted as the action.

In detail, consider $Q\times\mathbb{R}$, where $Q$ is the configuration space which is enlarged by an extra dimension. The Lagrangian is defined as a function \( L \colon T(Q\times \R) \) that is only zeroth-order on \(z\), that is, if \((q^i, z)\) is a local chart of \(Q\times\R\) and $(q^i,z,\dot{q}^i,\dot z)$ the corresponding local trivialisation of \(T(Q\times \R)\), then \(L\) does not depend on \(\dot{z}\) (or equivalently, \(dL\)  annihilates the vertical vector field \(\frac{\partial}{\partial \dot z}\)). The constraint one imposes is
\begin{equation}\label{eq:constMech}
\dot{z}=L(q^i,\dot{q}^i,z),
\end{equation}
so that for trajectories that satisfy the constraint we have
\begin{equation}
z(t) = z(0) + \int_0^t L(q(t), \dot{q}(t), z(t)) \ud t,
\end{equation}
and indeed \(z\) tracks the action along the path, as claimed.

Let $\Omega(I,q_a,q_b,s_a)$ be the set of curves $(q,z):I=[a,b]\rightarrow Q\times\mathbb{R}$ such that $q(a)=q_a$, $q(b)=q_b$, $z(a)=s_a$. The Herglotz problem is to determine the extrema of the functional
\begin{equation*}
    \begin{aligned}
        S: \Omega(I,q_a,q_b,s_a) &\longrightarrow \mathbb{R}\\
        (q,z) &\longmapsto z(a) - z(b),
    \end{aligned}
\end{equation*}
subject to \cref{eq:constMech}. For trajectories that satisy the constraint we have
\begin{equation}
    S(q,z) = z(b) - z(a) = \int_a^b L(q(t), \dot{q}(t), z(t)) \ud t,
\end{equation}
which resembles the classical expression of the action.

This optimization problem can be formulated equivalently using the method of Lagrange multipliers \cite{de_leon_constrained_2021}. Consider the Lagrangian
$$
 \tilde{L}(q,z,\dot q,\dot z)=\dot{z}-\lambda(\dot{z}-L(q,z,\dot q))\,,
$$
where $\lambda$ is the Lagrange multiplier. Then the extrema of \(S\) subject to \cref{eq:constMech} will be the unconstrained extrema of:
\begin{equation}
    \begin{aligned}
        \tilde{S}: \Omega(I,q_a,q_b,s_a) &\longrightarrow \mathbb{R}\\
        (q,z) &\longmapsto \int_a^b \tilde{L}(q(t),\dot{q}(t), z(t),\dot{z}(t)) \ud t\,.
    \end{aligned}
\end{equation}

Because we are looking for unconstrained extrema, they will be the solutions of the Euler-Lagrange equations for \(\tilde{L}\). The equation for \(z\) is
\begin{equation*}
	0 = \frac{\partial \tilde{L}}{\partial z} - \frac{\d}{\d t} \frac{\partial
	\tilde{L}}{\partial \dot{z}} = \lambda \frac{\partial L}{\partial z} + \dot{\lambda},
\end{equation*}
or equivalently
\begin{equation} \label{eq:euler-lagrange multiplier}
	\dot{\lambda} = -\lambda \frac{\partial L}{\partial z}. 
\end{equation}
The Euler-Lagrange equations for the other coordinates are
\begin{equation*}
	0 = \frac{\partial \tilde{L}}{\partial q^i} - \frac{\d}{\d t} \frac{\partial
	\tilde{L}}{\partial \dot{q}^i} = \lambda \frac{\partial L}{\partial q^i} -
	\dot{\lambda}\frac{\partial L}{\partial \dot{q}^i} - \lambda \frac{\d}{\d
	t}\frac{\partial L}{\partial \dot{q}^i}\,,
\end{equation*}
and after substituting in \cref{eq:euler-lagrange multiplier} and dividing through by \(
\lambda \) one finds
\begin{equation} \label{eq:lagrange multiplier action}
	0 = \frac{\partial L}{\partial q^i} - \frac{\d}{\d t} \frac{\partial L}{\partial
	\dot{q}^i} + \frac{\partial L}{\partial z}\frac{\partial L}{\partial \dot{q}^i}\,.
\end{equation}
These are the Herglotz equations.

\subsection{Action-dependent field theory}
We now introduce the Herglotz problem for field theories and derive the corresponding Herglotz equations.

\subsubsection{Classical field theory and Lagrangian densities}\label{sec:lagrangian
densities}
The passage from mechanics to field theory requires some care. The given data is usually some smooth fiber bundle \( E \) over a base \( M \) of dimension \(n\), which we assume to be orientable and hence endowed with at least a volume form. The base \(M\) is usually, but not always, taken to represent spacetime. Field configurations are sections of this bundle, and the values that the field takes are modelled by the fiber of \( E \). The basic problem is to identify the configurations that are extrema of a given action functional \( S \colon \Gamma(E) \to \R \). The action is usually written as the integral of a Lagrangian over a region \(D \subseteq M\) of the base, i.e. a Lagrangian is some sort of map from field configurations to top forms of $M$, \( \L \colon \Gamma(E) \to \Omega^{n}(M) \), such that
\begin{equation}
    S(\phi) = \int_D \L(\phi).
\end{equation}
One of the fundamental constraints on \( \L \) is that it be local, i.e. \( \L(\phi)_p \) should only depend on the value of the field \(\phi\) at \(p\) a finite number of its derivatives at \(p\). In other words, \( \L \) is to be a bundle map from the \( k \)-th jet bundle of \(E\), \(J^kE\), to the bundle of top forms \( \bigwedge^{n}T^\ast M\), in such a way that if \(j^k \phi \in \Gamma(J^kE) \) is the prolongation of some field configuration \(\phi \in \Gamma(E) \) then
\begin{equation}
    S(\phi) = \int_D \L \circ j^k\phi.
\end{equation}
The integer \(k\) is called the order of the Lagrangian. Volume forms are one-dimensional, which means that for a given choice of coordinates of the base, \( x^\mu\) then there exists a unique \(L\colon J^1E \to \R\) such that, on the coordinate domain,
\begin{equation}
     \L \circ j^1\phi = (L\circ j^1\phi) \ud^n x,
\end{equation}
where \( \ud^nx \) is the local volume form of \(M  \) induced by the coordinates.

Using the calculus of variations one can show that the stationary configurations of an action functional defined by a first order Lagrangian satisfy the Euler-Lagrange equations of field theory
\begin{equation*}
	\frac{\partial L}{\partial \phi^a} - \partial_\mu \frac{\partial L}{\partial
	\phi^a_\mu} = 0. 
\end{equation*}
These expression make sense given the choice of a local trivialisation of \(E\), \((x^\mu, \phi^a)\), which gives rise to a local trivialisation of \(J^1E\), \((x^\mu, \phi^a, \phi^a_\mu)\). Note that the Einstein summation convention is assumed from this point on, unless otherwise stated.

\subsubsection{The action flux}
We now wish to generalise this description to account for action dependent Lagrangians. A cursory look at the Herglotz equations would suggest a field theoretic analog of the form 
\begin{equation}
	\frac{\partial L}{\partial \phi^a} - \partial_\mu \frac{\partial L}{\partial
\phi ^a_\mu} + \frac{\partial L}{\partial \phi^a_\mu}\frac{\partial
L}{\partial z^\mu} = 0. 
\end{equation}
This equations are the ones also proposed in the literature \cite{Gaset2020a,Lazo2018,de_leon_multicontact_2022}. The question is then what should the geometric nature of \(z\) be. In the language of bundles we have just introduced, the constraint in \cref{eq:constMech} becomes
\begin{equation}\label{eq:constraint field theory integral}
    z(b) - z(a) = \int_{[a,b]}  \L \circ j^kq.
\end{equation}
The field theoretic version should then be 
\begin{equation}
    \int_{\partial D}z = \int_D  \L \circ j^k\phi,
\end{equation}
where $D\subset M$ is an $n$-dimensional submanifold of \(M\) over which we wish to extremise the action. It is now clear that \(z\) must be a form of degree \(n-1\) so that it can be integrated over submanifolds of the base \(M\) of codimension 1. In other words, \(z\) is the action flux. The differential version of \cref{eq:constraint field theory integral} is then
\begin{equation} \label{eq:constraint field theory}
\ud z =  \L \circ j^k\phi.
\end{equation}
analogous to \cref{eq:constMech}. 

There is, by way of contraction with a volume form, an isomorphism between \((n-1)\)-forms and vector fields, such that the exterior derivative becomes the divergence. In particular, a choice of coordinates \(x^\mu\) on the base induces a trivialisation \((x^\mu,z^\nu)\) on \(\bigwedge^{n-1} T^\ast M\), in such a way that for \(\alpha \in \bigwedge T^\ast _p M\)
\begin{equation}\label{eq:coordinates flux}
\alpha_p = z^\nu(\alpha_p) \frac{\partial}{\partial x^\nu} \lrcorner \d^n x,
\end{equation}
where the symbol \(\lrcorner \) denotes the contraction of a tangent vector with a form. Then, for a differential form, \(z\), of degree \(n-1\) whose components in these coordinates are \(z^\nu\), it holds that
\begin{equation}
    \d z = \partial_\nu z^\nu \d^nx.
\end{equation}
This means that the coordinate expression of \cref{eq:constraint field theory} is 
\begin{equation} \label{eq:constraint field theory coordinates}
	\partial_\nu z^\nu = L(\phi^a, \phi^a_\mu).
\end{equation}

For some bundle \(E \to M\) the corresponding Herglotz problem is formulated in the enlarged bundle \(E \oplus \bigwedge^{n-1}T^\ast M \to M \), where \(\oplus\) is the Whitney sum. Consider a Lagrangian of the form \(\L \colon J^kE \oplus \bigwedge^{n-1}T^\ast M \to \bigwedge^n T^\ast M \) (so that crucially \(\L\) does not depend on any of the derivatives of the action flux). The Herglotz problem for field theory is then to find the sections \((\phi,z)\) that extremise the functional \(S(\phi,z) = \int_{\partial  D} z\) subject to the constraint \(\ud z = \L\). If \((\phi, z)\) is one such section then
\begin{equation*}
	S(\phi,z) = \int_{\partial D} z = \int_D \ud z = \int_D  \L \circ (j^k \phi, z),
\end{equation*}
and we can interpret \(S\) as the action. 

\subsubsection{Constrained optimisation in field theory}
Just like before, we turn this constrained optimisation problem into an
unconstrained one using Lagrange multipliers. The expanded action for a first order Lagrangian, in analogy with
\cref{eq:lagrange multiplier action}, is
\begin{equation}\label{eq:expanded action field theory}
	\tilde{S}(\phi, z) = \int_D \big[(1 - \lambda) \ud z + \lambda \L \circ (j^1\phi,z)\big]
	= \int_D \ud^n x \big[ (1 - \lambda) \partial_\mu z^\mu + \lambda L(\phi^a,
	\partial_\mu \phi^a, z^\nu) \big].
\end{equation}
Let us write down the integrand of
\cref{eq:expanded action field theory} as an expanded Lagrangian:
\begin{equation} \label{eq:expanded lagrangian field theory}
	\tilde{\L}\circ(j^1\phi, j^1 z) = \tilde{L}(\phi^a, \partial_\mu \phi^a, z^\nu, \partial_\mu z^\nu)\ud^n x = \big[ (1 - \lambda)
	\partial_\mu z^\mu + \lambda L(\phi^a, \partial_\mu \phi^a, z^\nu) \big] \ud^n x.
\end{equation}
Note that \( \Tilde{\L} \) is now the Lagrangian for a theory defined on the expanded bundle \(J^1E \oplus J^1\bigwedge^{n-1}T^\ast M \to M\), so \( z\) is a dynamical degree of freedom. 

Given that the Lagrangian is of first order, extrema of this action functional will be
solutions to the Euler-Lagrange equations for this Lagrangian, which become the Herglotz
equations upon imposing the constraint. We write them down in the next section.
Nevertheless there is nothing preventing one from calculating the explicit variation of
the action, which leads to the equations of motion for a theory of any order. This is the
approach we follow in the next section.

\subsubsection{The Herglotz equations for field theory}
Finally, we derive the Herglotz equations for field theory from the expanded
Lagrangian  in \cref{eq:expanded lagrangian field theory}. The
equations for the action flux are
\begin{equation*}
	0 = \frac{\partial \tilde{L}}{\partial z^\nu} - \partial_\mu \frac{\partial
	\tilde{L}}{\partial z^\nu_\mu} = \lambda \frac{\partial L}{\partial z^\nu} +
	\partial_\mu(\lambda \delta_\nu^\mu) = \lambda \frac{\partial L}{\partial z^\nu} +
	\partial_\nu \lambda,
\end{equation*}
where \((x_\mu, z^\nu, z^\nu_\mu)\) is the trivialisation of \(J^1(\bigwedge^{n-1}T^\ast M) \) induced by the choice of coordinates on the base, as defined by \cref{eq:coordinates flux}.
Rearranging, one obtains
\begin{equation} \label{eq:euler lagrange multiplier field theory}
	\partial_\nu \lambda = - \lambda \frac{\partial L}{\partial z^\nu}. 
\end{equation}
This equation actually constrains the type of action dependence that is allowed
in \( L \). We will see later on that in the context of relativity it forces the
dissipation form to be closed. 

The equations for the field are
\begin{equation*}
	0 = \frac{\partial \tilde{L}}{\partial \phi^a} - \partial_\mu \frac{\partial
	\tilde{L}}{\partial\phi^a_\mu} = \lambda \frac{\partial L}{\partial \phi^a} -
	(\partial_\mu \lambda) \frac{\partial L}{\partial\phi^a_\mu} - \lambda
	\partial_\mu \frac{\partial L}{\partial\phi^a_\mu},
\end{equation*}
and, after substituting in \cref{eq:euler lagrange multiplier field theory} and dividing
through by \( \lambda \), we arrive at the field theoretical Herglotz equations
\begin{equation} \label{eq:herglotz field theory}
	\frac{\partial L}{\partial \phi^a} - \partial_\mu \frac{\partial
	L}{\partial\phi^a_\mu} + \frac{\partial L}{\partial z^\mu} \frac{\partial
L}{\partial\phi^a_\mu} = 0. 
\end{equation}

\section{Action-dependent Einstein gravity}\label{ch:einstein}

In this section we apply the language and tools developed in the previous section to the
specific case of Einstein gravity. We first describe the Lagrangian from which the Einstein
field equations come and then introduce an action-dependent version of it and
derive its field equations. 

\subsection{The Einstein-Hilbert Lagrangian}
As is well-known, the Einstein field equations can be obtained from a variational
principle. The classical Lagrangian that gives rise to these equations is the Einstein-Hilbert
Lagrangian. We formulate it in the language of \cref{sec:lagrangian densities}. The field of interest in relativity is the metric on the given spacetime \(M\) ---which we take to be of dimension 4 and orientable from now on---. Hence the bundle of interest to us is a subbundle of the second symmetric power of the cotangent bundle of \( M\), 
\begin{equation}
    S^2 T^\ast M \to M.
\end{equation}
Specifically it is the subbundle determined by the condition of non-degeneracy. We denote it by \( G(M) \to M \).  

As advertised, the theory of general relativity is a second-order theory, which means that the Einstein-Hilbert Lagrangian must be a bundle map from \( J^2 G(M)\) to \(\bigwedge^n T^\ast M \). Specifically, given a metric \( g \in \Gamma(G(M)) \),
\begin{equation} \label{eq:EH lagrangian}
     \L_\text{E-H} \circ j^2 g = R(g) \omega_g.
\end{equation}
Here we use \( \omega_ g \) to denote the volume form determined by \(g\), which in a choice of coordinates  \(x^\mu\) becomes
\begin{equation}
    \sqrt{g} \d^4 x,
\end{equation}
where \( \sqrt{g} \) is the square root of the absolute value of the determinant of the expression of the metric in the coordinates \(x^\mu\). 
The other factor, \(R(g)\), is the scalar curvature of \(g\), which is defined as the trace of the Ricci tensor:
\begin{equation}
    R(g) = \tr(g^{-1} \Ric(g)).
\end{equation}
The Ricci tensor is of type \((0,2) \), so by contracting with \(g^{-1}\), the metric induced on \(T^\ast M\), we obtain a \((1,1)\) tensor, whose trace is well-defined. The coordinate expression of the components of the Ricci tensor, \(R_{ab}\), is
\begin{equation} \label{eq:ricci tensor}
	R_{ab} = \partial_m{\Gamma^m}_{ab} - \partial_a {\Gamma^m}_{mb} + {\Gamma^m}_{mn}
	{\Gamma^n}_{ab} - {\Gamma^m}_{an}{\Gamma^n}_{mb},
\end{equation}
where \( \Gamma_{ab}^c \) are the Christoffel symbols of the Levi-Civita connection determined by \(g\). These contain first derivatives of the metric, so \( R_{ab} \) and
hence \( R \) contain second derivatives of the metric, and the Einstein-Hilbert Lagrangian is indeed of second order.

The Einstein-Hilbert action is therefore
\begin{equation} \label{eq:EH action}
	S_{\text{E-H}}(g) = \int_D \L_\text{E-H} \circ j^2g = \int_D R
	\sqrt{g}\ud^4x,
\end{equation}
for some domain \(D\) on which the integral is finite. 
A variation of this action leads one to the Einstein field equations 
\begin{equation} \label{eq:EFE vacuum}
	R_{ab} - \tfrac{1}{2}g_{ab}R = 0. 
\end{equation}
More precisely, these are the Einstein field equations in a vacuum, since one can add
various matter terms to the Einstein-Hilbert Lagrangian which lead to the Einstein
equations in the presence of matter,
\begin{equation} \label{eq:EFE matter}
	R_{ab} - \tfrac{1}{2}g_{ab}R = T_{ab}. 
\end{equation}
The object \( T_{ab} \) is the energy-momentum tensor and collects all of the terms coming
from the presence of matter. See \S4 of \cite{Carroll1997} for a detailed derivation.

\subsection{An action dependent Einstein-Hilbert Lagrangian}
What kind of action dependence can we incorporate into the Einstein-Hilbert Lagrangian?
The simplest one is a linear dissipation term:
\begin{equation} \label{eq:action dependent einstein-hilbert}
	\L_{\text{E-H}} \circ (j^2 g,z) = R\omega_g - \theta \wedge z. 
\end{equation}
Now \( \L_{\text{E-H}} \) is defined on the expanded bundle, \( G(M) \oplus \bigwedge^3T^\ast M \to M \). Hence \( \theta \) must be a 1-form on \(M\), which we will refer to as the \emph{dissipation form}.

The coordinate expression of this dissipation term is
\begin{equation*}
	\theta \wedge z = (\theta_\mu \ud x^\mu) \wedge \left(z^\nu \frac{\partial}{\partial x^\nu}\lrcorner \ud^4 x\right) = \theta_\mu z^\nu \ud
	x^\mu \wedge \left(\frac{\partial}{\partial x^\nu}\lrcorner \ud^4 x\right) = \theta_\mu z^\mu \ud^4x
\end{equation*}
where once again we use the coordinates defined in \cref{eq:coordinates flux}. Then \cref{eq:action dependent einstein-hilbert} becomes 
\begin{equation}\label{eq:action dependent EH coordinates 1}
	\L_{\text{E-H}} \circ (j^2g, z) = (R\sqrt{g} - \theta_\mu
	z^\mu) \ud^4 x. 
\end{equation}
This Lagrangian does not exactly match the one proposed in Equation (9) of \cite{Lazo2017}. The discrepancy is down to a different choice of coordinates. Indeed, in the previous computation we used the isomorphism between \(\Omega^3(M)\) and \(\Gamma(TM)\) induced by contracting with \(\d^4x\). Instead we may contract with the volume form induced by the metric, \(\omega_g\). Let \(\zeta^\mu\) be the components of \(z\) in this new choice of coordinates, i.e.
\begin{equation*}
	z = \zeta^\mu \frac{\partial}{\partial x^\mu} \lrcorner \omega_{g} = \zeta^\mu \sqrt{g} \frac{\partial}{\partial x^\mu} \lrcorner \ud^4x,
\end{equation*}
which implies \( z^\mu = \sqrt{g} \zeta^\mu \). In these new coordinates \cref{eq:action dependent EH coordinates 1} looks like
\begin{equation}\label{eq:action dependent EH coordinates 2}
	\L_{\text{E-H}} \circ (j^2g, z) = (R\sqrt{g} - \theta_\mu
	\zeta^\mu \sqrt{g}) \ud^4 x = (R - \theta_\mu \zeta^\mu)\omega_g.
\end{equation}
This is the Lagrangian proposed in Equation (9) of \cite{Lazo2017}. 

We now write down the constraint in \cref{eq:constraint field theory} for this Lagrangian.
In the original coordinates for the action flux we have
\begin{equation*}
	\ud z = \partial_\mu z^\mu \ud^4 x,
\end{equation*}
so
\begin{equation} \label{eq:constraint coordinates 1}
	\partial_\mu z^\mu = R\sqrt{g} - \theta_\mu z^\mu. 
\end{equation}
In the other set of coordinates, induced by contracting with \(\omega_g\), one sees
\begin{equation*}
	\ud z = \partial_\mu (\sqrt{g} \zeta^\mu) \ud^4 x = \nabla_\mu \zeta^\mu \sqrt{g}\ud^4x
	= \nabla_\mu \zeta^\mu \omega_g,
\end{equation*}
where \( \nabla \) is the covariant derivative induced by \( g \). We have made use of a
useful identity about the divergence:
\begin{equation} \label{eq:divergence identity}
	\nabla_\mu X^\mu = \frac{1}{\sqrt{g}} \partial_\mu (\sqrt{g}X^\mu).
\end{equation}
This is the statement that the divergence induced by the volume form of a metric coincides with the trace of the covariant derivative. 

In the new coordinates the constraint becomes
\begin{equation} \label{eq:constraint coordinates 2}
	\nabla_\mu \zeta^\mu = R - \theta_\mu \zeta^\mu,
\end{equation}
which is the same form that appears in Equation (8) of \cite{Lazo2017}. 

\subsection{Derivation of the field equations}
We now apply the method of Lagrange multipliers, as
described in previous section, to derive a modified version of Einstein's equations. The
expanded Lagrangian is
\begin{equation*}
\tilde{\L}_\text{E-H} \circ (j^2 g, j^1 z) = \big[(1-\lambda) \partial_\mu z^\mu + \lambda(R\sqrt{g} - \theta_\mu z^\mu)\big] \ud^4
	x.
\end{equation*}
We will compute the variation of the corresponding expanded action, \(\Tilde{S}(g,z) = \int_D \Tilde{L}_\text{E-H} \circ (j^2g,j^1z)\), with respect to the two dynamical degrees of freedom, \(z\) and \(g\).

\subsubsection{Variation of the action flux}\label{sec:variationonaction}
The variation with respect to the action flux is
\begin{align}
	\delta \tilde{S}(g, z) & = \int_D \big[(1-\lambda) \delta \partial_\mu z^\mu +
\lambda(\delta(R\sqrt{g}) - \theta_\mu \delta z^\mu)\big] \ud^4 x \notag \\						& = \int_D (1 - \lambda) \partial_\mu \delta z^\mu -
			\lambda \theta_\mu \delta z^\mu \ud^4x + \int_D \lambda
			\delta(R \sqrt{g}) \ud^4 x \notag \\
& = \int_{D} \partial_\mu \big((1 -
		\lambda)\delta z^\mu\big) \ud^4 x + \int_D (\partial_\mu \lambda
		- \lambda\theta_\mu) \delta z^\mu \ud^4 x + \int_D \lambda
		\delta(R \sqrt{g}) \ud^4 x \label{eq:variation expanded
										action}. 
\end{align}
The first integral is a boundary term coming from an integration by parts. It vanishes
if we assume the variations vanish at the boundary of \( D \). If the action is stationary then its variation must vanish for any variation of the fields. This means that the second term of \cref{eq:variation expanded action} must vanish, since in particular we may choose not to vary the metric. Hence, the quantity inside the brackets must vanish, since it vanishes when integrated against any variation. Therefore
\begin{equation} \label{eq:action flux variation}
	\partial_\mu \lambda = \lambda\theta_\mu.
\end{equation}
In other words, \( \ud \lambda = \lambda \theta \). As we had advertised before, this forces the dissipation form \(\theta\) to be closed, as
\begin{equation*}
	\ud(\lambda \theta) = \ud \lambda \wedge \theta + \lambda \ud \theta = \lambda \theta
	\wedge \theta + \lambda \ud \theta = \lambda \ud \theta,
\end{equation*}
hence
\begin{equation*}
	\lambda \ud \theta = \ud(\lambda \theta) = \ud^2 \lambda = 0,
\end{equation*}
and we conclude \( \ud \theta = 0 \) provided \(\lambda\) does not vanish. 

\subsubsection{Variation of the metric}
We retake the calculation from \cref{eq:variation expanded action}. We may now only consider the last integral, as we can vary \(g\) and \(z\) independently. We will follow the
derivation in \cite{Carroll1997} for as long as we can. In particular, we take the spacetime \(M\) to be closed, and hence avoid consideration of Gibbons-Hawking-York type boundary terms. Since \( R\sqrt{g} =
g^{ab}R_{ab}\sqrt{g} \), from the product rule its variation results in three terms:
\begin{equation}\label{eq:variation of scalar curvature}
	\int_D \lambda \delta(R \sqrt{g}) \ud^4 x = \int_D \lambda \delta g^{ab} R_{ab} \sqrt{g}
	\ud^4 x + \int_D \lambda g^{ab} \delta R_{ab} \sqrt{g} \ud^4 x + \int_{D} \lambda R
	\delta\sqrt{g} \ud^4 x
\end{equation}
The first term is already in the form required to apply the fundamental theorem of the
calculus of variations. For the third one uses the standard result
\begin{equation*}
	\delta \sqrt{g} = -\tfrac{1}{2}\sqrt{g} g_{ab} \delta g^{ab}.
\end{equation*}
The first and third terms of \cref{eq:variation of scalar curvature} can be combined into
\begin{equation}\label{eq:variation vacuum}
	\int_D \lambda (R_{ab} - \tfrac{1}{2} Rg_{ab}) \delta g^{ab} \sqrt{g} \ud^4 x.
\end{equation}
In the standard derivation of Einstein's equations, one shows that the middle integral of
\cref{eq:variation of scalar curvature} actually vanishes, so that if \cref{eq:variation
vacuum} is to vanish for any variation \( \delta g_{ab} \), or equivalently for any
variation of the inverse metric \( \delta g^{ab} \), the integrand of \cref{eq:variation vacuum} itself must
vanish. This gives Einstein's equations. In the presence of \( \lambda \), however, the
middle integral does not vanish and contributes additional terms to the
equations.

We compute the variation of the middle integral in \cref{eq:variation of scalar curvature}.  The variation of the Ricci curvature can be shown to be
\begin{equation}\label{eq:variation ricci curvature}
	g^{ab} \delta R_{ab} = g^{ab}(\nabla_m \delta{\Gamma^m}_{ab} - \nabla_a \delta
	{\Gamma^m}_{mb}) = \nabla_n(g^{ab} {\delta\Gamma^n}_{ab} - g^{nb} \delta
	{\Gamma^m}_{mb})\,,
\end{equation}
so
\begin{equation*}
	\int_D \lambda g^{ab}\delta R_{ab} \sqrt{g} \ud^4 x = \int_D \lambda \nabla_n(g^{ab}
	{\delta\Gamma^n}_{ab} - g^{nb} \delta {\Gamma^m}_{mb}) \sqrt{g} \ud^4x,
\end{equation*}
and if \( \lambda \) weren't there this integral would vanish because of the divergence
theorem and the fact that the variations vanish on the boundary of \( D \). In the
presence of \( \lambda \) we perform an integration by parts:
\begin{align*}
	& \int_D \lambda g^{ab}\delta R_{ab} \sqrt{g} \ud^4 x = \\
	& \quad = \int_D \lambda \nabla_n(g^{ab} {\delta\Gamma^n}_{ab} - g^{nb} \delta
	{\Gamma^m}_{mb}) \sqrt{g} \ud^4x \\
	& \quad = \int_D \nabla_n \left(\lambda (g^{ab} {\delta\Gamma^n}_{ab} - g^{nb}
	\delta {\Gamma^m}_{mb})\right) \sqrt{g}\ud^4 x - \int_D (\nabla_n \lambda) (g^{ab}
	{\delta\Gamma^n}_{ab} - g^{nb} \delta {\Gamma^m}_{mb}) \sqrt{g} \ud^4 x. 
\end{align*}
The first integral vanishes because it is the integral of a divergence and the variations
vanish on the boundary of \( D \). The second integral is where the additional terms will
come from. We split it into two terms. 

The variation of the Christoffel symbols can be shown to be
\begin{equation} \label{eq:variation christoffel symbols}
	\delta {\Gamma^a}_{bc} = \tfrac{1}{2} g^{am}(\nabla_c \delta g_{bm} + \nabla_b \delta
	g_{mc} - \nabla_m \delta g_{bc}).
\end{equation}
Using this and \cref{eq:action flux variation} (since \( \nabla_n \lambda = \partial_n
\lambda \)) we compute for the first integral
\begin{equation}\label{eq:first three terms}
	-\int_D (\nabla_n \lambda) g^{ab} \delta{\Gamma^n}_{ab} \sqrt{g} \ud^4 x = -
	\tfrac{1}{2} \int_D \lambda\theta_n g^{ab} g^{nk}(\nabla_b \delta g_{ak} + \nabla_a
	\delta g_{kb} - \nabla_k \delta g_{ab}) \sqrt{g} \ud^4 x. 
\end{equation}
The presence of \( g^{ab} \) means the indices \( a \) and \( b \) are symmetrised, so
\begin{equation*}
	g^{ab} \nabla_b \delta g_{ak} = g^{ab} \nabla_a \delta g_{kb}. 
\end{equation*}
This means \cref{eq:first three terms} simplifies to
\begin{align}
	&	-\int_D (\nabla_n \lambda) g^{ab} \delta{\Gamma^n}_{ab} \sqrt{g} \ud^4 x = \notag \\
	& \quad = - \int_D \lambda \theta_n g^{ab}g^{nk} \nabla_b \delta g_{ak} \sqrt{g} \ud^4 x
	+ \tfrac{1}{2} \int_D \lambda \theta_n g^{ab}g^{nk} \nabla_k \delta g_{ab} \sqrt{g}
	\ud^4 x \notag \\
	& \quad = - \int_D \lambda \theta_n \nabla_b (g^{ab}g^{nk} \delta g_{ak}) \sqrt{g} \ud^4 x
	+ \tfrac{1}{2} \int_D \lambda \theta_n \nabla_k (g^{ab}g^{nk}\delta g_{ab}) \sqrt{g}
	\ud^4 x. \label{eq:two integrals}
\end{align}
Let's perform an integration by parts for the first integral. Introducing the shorthand \( X^{bn} = g^{ab}g^{nk}\delta g_{ak} \), we compute
\begin{equation*}
	\nabla_c(\lambda\theta_n X^{bn}) = \nabla_c(\lambda \theta_n)X^{bn} + \lambda \theta_n
	\nabla_c X^{bn},
\end{equation*}
so
\begin{align*}
	-\int_D \lambda \theta_n \nabla_b (g^{ab}g^{nk} \delta g_{ak})\sqrt{g}\ud^4x 
	& = - \int_D \lambda \theta_n \nabla_bX^{bn} \sqrt{g} \ud^4 x \\
	& = - \int_D \nabla_b(\lambda \theta_n X^{bn})\sqrt{g}\ud^4x + \int_{D}
	\nabla_b(\lambda\theta_n)X^{bn}\sqrt{g}\ud^4x. 
\end{align*}
The first integral is the integral of a divergence, so it vanishes. We are left with the
second which we can expand into
\begin{align*}
	\int_D \nabla_b(\lambda \theta_n) X^{bn}\sqrt{g}\ud^4 x 
	& = \int_D (\theta_n \partial_b \lambda + \lambda \nabla_b\theta_n)(g^{ab}g^{nk}\delta
	g_{ak})\sqrt{g} \ud^4 x \\
	& = \int_D \lambda(\theta_b\theta_n +
	\nabla_b\theta_n)(g^{ab}g^{nk}\delta g_{ak})\sqrt{g}\ud^4 x.
\end{align*}
As a last step, we use the identity
\begin{equation*}
	\delta g^{ab} = - g^{am}g^{bn} \delta g_{mn}
\end{equation*}
to write our integral as a variation with respect to the inverse metric.
\begin{align*}
	\int_D \lambda(\theta_b\theta_n + \nabla_b\theta_n)(g^{ab}g^{nk}\delta
	g_{ak})\sqrt{g}\ud^4 x 
	& = -\int_D \lambda(\theta_b\theta_n + \nabla_b \theta_n)\delta g^{bn} \sqrt{g} \ud^4 x. 
\end{align*}

Without going through the details again, the other integral in \cref{eq:two integrals} can
be brought to the form
\begin{align*} 
	\tfrac{1}{2} \int_D \lambda \theta_n \nabla_k (g^{ab}g^{nk}\delta g_{ab}) \sqrt{g} \ud^4
	x 
	& = - \tfrac{1}{2} \int_D \nabla_k(\lambda \theta_n) g^{ab}g^{nk}\delta g_{ab} \sqrt{g}
	\ud^4 x \\
	& = \tfrac{1}{2} \int_D \lambda(\theta_k\theta_n + \nabla_k\theta_n) g^{ab}g^{nk}
	g_{ma}g_{lb}\delta g^{ml} \sqrt{g}\ud^4 x \\
	& = \tfrac{1}{2}\int_D \lambda g^{nk}(\theta_k\theta_n + \nabla_k \theta_n)g_{ml} \delta
	g^{ml} \sqrt{g} \ud^4 x.
\end{align*}

There is still another integral we need to evaluate, the second term in the variation of
\( R_{ab} \), namely
\begin{align}
	\int_D (\partial_n \lambda) g^{nb}\delta{\Gamma^m}_{mb} \sqrt{g} \ud^4 x 
	& = \tfrac{1}{2} \int_D \lambda \theta_n g^{nb} g^{mk}(\nabla_b \delta g_{mk} + \nabla_m
	\delta g_{kb} - \nabla_k \delta g_{mb}) \sqrt{g} \ud^4 x. 
\end{align}
Because \( m \) and \( k \) are symmetrised, the second and third terms cancel, leaving
us with
\begin{align}
	\tfrac{1}{2}\int_D \lambda \theta_n g^{nb}g^{mk}\nabla_b \delta g_{mk} \sqrt{g} \ud^4 x
	& = - \tfrac{1}{2} \int_D \nabla_b(\lambda \theta_n) g^{nb}g^{mk} \delta g_{mk} \sqrt{g}
	\ud^4 x \\
	& = \tfrac{1}{2} \int_D \lambda(\theta_b\theta_n + \nabla_b
	\theta_n)g^{nb}g^{mk}g_{am}g_{lk}\delta g^{al} \sqrt{g} \ud^4 x \\
	& = \tfrac{1}{2} \int_D \lambda g^{nb}(\theta_b \theta_n + \nabla_b \theta_n)
	g_{al} \delta g^{al} \sqrt{g} \ud^4 x. 
\end{align}

We have calculated all the integrals we need. Before we put them all together, let us make
the following observation:
\begin{equation*}
	\nabla_a \theta_b = \partial_a \theta_b - {\Gamma^m}_{ab} \theta_m = \partial_b \theta_a
	- {\Gamma^m}_{ba} \theta_m = \nabla_b \theta_a,
\end{equation*}
which uses the fact that \( \theta \) must be closed. We may therefore define the following (0,2) symmetric tensor
\begin{equation} \label{eq:K}
\mathbf{K} = \theta \otimes \theta + \nabla \theta,
\end{equation}
whose components are
\begin{equation}
	K_{ab} = \theta_a\theta_b + \frac12\left(\nabla_{a}\theta_{b}+\nabla_{b}\theta_{a}\right) = \theta_a\theta_b + \nabla_{(a}\theta_{b)}=\theta_a\theta_b + \nabla_{a}\theta_{b},
\end{equation}
where parentheses surrounding indices indicate symmetrisation. 
All three expressions are equal because $\nabla_a \theta_b= \nabla_b \theta_a$. Nevertheless, we will use the second one to make the symmetry of the indices explicit. So, after liberal relabeling of indices, we find that \cref{eq:variation expanded action}
becomes
\begin{equation} \label{eq:final variation of action}
	\delta\tilde{S}[g_{ab}, z^\mu] = \int_D(\partial_\mu \lambda - \lambda \theta_\mu)
	\delta z^\mu \ud^4 x + \int_D \lambda (R_{ab} - \tfrac{1}{2} Rg_{ab} - K_{ab} +
	Kg_{ab})\delta g^{ab} \sqrt{g} \ud^4 x,
\end{equation}
with \( K_{ab} \) defined as in \cref{eq:K} and \( K = g^{mn}K_{mn} \) its trace. 

Applying the fundamental theorem of the calculus of variations, the action will be
stationary if and only if the integrands of both terms vanish. From the first integral we
get \cref{eq:action flux variation}, which we have already used. And from the second one
we get the modified Einstein field equations
\begin{equation} \label{eq:modified EFE}
	R_{ab} - \tfrac{1}{2}Rg_{ab} - K_{ab} + Kg_{ab} = 0.
\end{equation}

These equations coincide with the ones derived in \cite{Lazo2021}.

\section{Significance of the equations}\label{ch:significance}

In this section we discuss the equations we have obtained and how they compare to those
appearing in existing publications. We also make the case that the version we have derived
is a more adequate version. 

\subsection{The dissipation tensor}
Let us recap what we did in the previous section. We have shown, by computing the
variation of the corresponding action, that the field equations of an Einstein-Hilbert
Lagrangian with linear dissipation, namely
\begin{equation} 
	L(g_{ab}, \partial_\mu g_{ab}, \partial_\mu \partial_\nu g_{ab}, z^\mu) = R(g_{ab}, \partial_\mu g_{ab}, \partial_\mu \partial_\nu g_{ab})
	\sqrt{g} - \theta_\mu z^\mu,
\end{equation}
are
\begin{equation}\label{eq:EFE with dissipation}
   R_{ab} - \tfrac{1}{2}Rg_{ab} - K_{ab} + Kg_{ab} = 0,
\end{equation}
where \( K_{ab} \) are the components of the \( (0,2) \) symmetric tensor
\begin{equation}
	\mathbf{K} = \nabla \theta + \theta \otimes \theta.
\end{equation}
We will call \( K \) the dissipation tensor. Since the first two terms of \cref{eq:EFE with dissipation} have zero divergence (they are the components of the Einstein tensor), it must be the case that, on-shell, the divergence of the second two terms also vanishes. This imposes a constraint on the space of solutions to \cref{eq:modified EFE}, which depends on the dissipation form $\theta$. Namely,
\begin{equation}
\nabla_a\left(g^{ab}K_{bc}-n\delta^a_cK\right)=0
\end{equation}

This means, that if we add a matter term to the Lagrangian, the resulting  tensor in the field equations may not be conserved.

This means, that if we were to couple a matter term to \cref{eq:action dependent einstein-hilbert}, its energy-momentum tensor need not in general have zero divergence. Specifically, what must have zero divergence will be a combination of the energy-momentum tensors of the matter fields and terms containing the dissipation 1-form. Nevertheless, further investigation is required to determine the precise way in which the dissipation tensor governs the non-conservation of other quantities.

\subsection{Non-covariance of existing equations}
These equations are not the ones obtained in \cite{Lazo2017}. For the same Lagrangian, the
equations derived are
\begin{equation} \label{eq:EFE lazo}
	R_{ab} + \tilde{K}_{ab} - \tfrac{1}{2}g_{ab}(R + \tilde{K}) = 0,
\end{equation}
where $\tilde{K}=g^{ab}\tilde{K}_{ab}$, and $\tilde{K}_{ab}$ is 
\begin{equation} \label{eq:}
	\tilde{K}_{ab} = \theta_m {\Gamma^m}_{ab} - \tfrac{1}{2}\left(\theta_a {\Gamma^m}_{mb} +
	\theta_b {\Gamma^m}_{am}\right). 
\end{equation}
These cannot possibly represent the components of a tensor. Very explicitly,  for the flat Minkowski metric, their
expression in Cartesian coordinates is 0. If they represented the components of a tensor then they would also vanish for any other choice of coordinates for the flat metric. Nevertheless, in spherical coordinates one computes
\begin{align*}
	{\Gamma^r}_{\theta\theta} & = -r & {\Gamma^\theta}_{r\theta} & = \frac{1}{r} &
	{\Gamma^\phi}_{r\phi} & = \frac{1}{r} \\
	{\Gamma^r}_{\phi\phi} & = -r \sin{\theta}^2 & {\Gamma^\theta}_{\phi\phi} & =
	-\sin{\theta}\cos{\theta} & {\Gamma^\phi}_{\theta\phi} & = \frac{1}{\tan{\theta}}. 
\end{align*}
Which means, for example,
\begin{equation*}
	\tilde{K}_{tr} = 0 - \tfrac{1}{2}(\theta_t {\Gamma^m}_{mr} + 0) =
	-\frac{\theta_t}{2r},
\end{equation*}
which is certainly non-zero if \( \theta_t \) does not vanish. Hence the object derived in \cite{Lazo2017} is not coordinate independent so it
cannot possibly represent meaningful physics. 

There is another fact that points to the equations in \cite{Lazo2017} not being what one would expect as the Herglotz equations coming from a second-order action-dependent Lagrangian. The Herglotz equations for the harmonic oscillator with
linear dissipation lead to equations linear in the dissipation coefficient (\cite{Gaset2020b}). However, the Lagrangian for this system is first
order, whereas, as we had already discussed, the Einstein-Hilbert Lagrangian is actually
second order. There is a second order Lagrangian with linear
dissipation, called the damped Pais-Uhlenbeck oscillator, whose equations of motion are derived in \cite{Leon2021a}. These are in fact not linear in the dissipation coefficient, but rather quadratic. In
our case, the dissipation form plays the role of the dissipation coefficient, and indeed, \(\mathbf{K}\) is quadratic in it. The equations in \cite{Lazo2017} instead lack a quadratic term.

One can pinpoint the exact reason for the problems with \cref{eq:EFE lazo}. One of the
simplifying assumptions made in their derivation was to only consider certain terms of the Ricci curvature.
Specifically, the Ricci curvature consists of four terms. Two of them are contractions of
the Christoffel symbols with themselves, the other two are derivatives of the Christoffel
symbols. In the classical case, without dissipation, one can show that the second two terms are actually a divergence, so they do not contribute to the variation of the Einstein-Hilbert action and the resulting equations remain unchanged (see \cite{Gaset2018,Maria2015}). For an action-dependent theory, however, adding a divergence to the unexpanded Lagrangian does not lead, in general, to the same equations \cite{inverse,Lazo2021}.

\section{Conclusions}\label{ch:conclusions}

The are three main ideas presented in this article.

Firstly, we show how the Herglotz problem can be turned from a constrained optimisation problem to an unconstrained one by promoting the action dependence to a dynamic degree of freedom and using Lagrange multipliers to implement the non-holonomic constraint.

Secondly, we describe how the Einstein-Hilbert Lagrangian can be modified with an action dependence in a coordinate-independent manner. This allows one to derive a correct, Lorentz-invariant set of field equations that remedy the issues present in previous derivations. 

Finally, the computation performed constitutes an important example for the ongoing development of contact geometry and its applications, since it is a singular second-order field theory. Having a concrete example at hand will aid to understand these systems.

There are various avenues for future follow-up work. One can consider more general dissipation terms to add to the Einstein-Hilbert Lagrangian, and study their phenomenology. It will also be interesting to consider the boundary effects in the case of manifolds with boundary (the appropriate Gibbons-Hawking-York term). 

General relativity has several equivalent formulations \cite{gaset2019,Vey_2015}. It would be interesting to add dissipation to these formalisms and study their properties and relations.

Finally, the current tools in contact geometry fall short of completely describing this kind of Lagrangians. A more general geometric structure, akin to multisymplectic geometry, needs to be developed for more general action-dependent Lagrangians in order to describe relevant theories. In this line, the multicontact structure recently presented in \cite{de_leon_multicontact_2022} could be the adequate geometric framework for action-dependent gravity.

\setquotestyle{english}
\printbibliography

\end{document}